\documentclass[aps,preprint,superscriptaddress, showpacs, showkeys]{revtex4}

\usepackage{graphicx}

\newcommand{\be}{\begin{equation}}
\newcommand{\ee}{\end{equation}}
\newcommand{\bea}{\begin{eqnarray}}
\newcommand{\eea}{\end{eqnarray}}

\newcommand{\vfi}{\varphi}
\newcommand{\Vm}{V^-}

\newcommand{\Fiu}{\Phi^u}
\newcommand{\Fip}{\Phi^p}
\newcommand{\In}{{\cal I}}

\newcommand{\tf}{\tilde{f}}
\newcommand{\cV}{{\cal V}}
\newcommand{\bal}{\bar{l}}

\newcommand{\LL}{\frac{1}{2\pi}\,\int_{-\infty}^{\infty}dk\sum_{n=-\infty}^{\infty}}
\newcommand{\LLnof}{\int_{-\infty}^{\infty}dk\sum_{n=-\infty}^{\infty}}
\newcommand{\FF}{\frac{1}{2\pi}\,\int\limits_{0}^{2\pi}\int\limits_{-\infty}^{\infty}d\vfi\, dz}

\begin{document}

\title{\vspace{0cm}\large\bf
Electrostatic Patch Effect in Cylindrical Geometry. III. Torques}

\author{Valerio Ferroni}
\affiliation{ICRANet, Dept. of Phys., Univ.  `La Sapienza', Rome, Italy\\
{\rm current address}: W.W.Hansen Experimental Physics Laboratory,\\
Stanford University, Stanford, CA 94305-4085, USA}
\email{vferroni@stanford.edu}
\author{Alexander S. Silbergleit}
\affiliation{Gravity Probe B, W.W.Hansen Experimental Physics Laboratory,\\
Stanford University, Stanford, CA 94305-4085, USA}
\email{gleit@stanford.edu}

\date{\today}

\begin{abstract}
We continue to study the effect of uneven voltage distribution on two close cylindrical conductors with parallel axes started in our papers~\cite{VA} and \cite{VA2}, now to find the electrostatic torques. We calculate the electrostatic potential and energy to lowest order in the gap to cylinder radius ratio for an arbitrary relative rotation of the cylinders about their symmetry axis. By energy conservation, the axial torque, independent of the uniform voltage difference, is found as a derivative of the energy in the rotation angle. We also derive both the axial and slanting torques by the surface integration method: the torque vector is the integral over the cylinder surface of the cross product of the electrostatic force on a surface element and its position vector. The slanting torque consists of two parts: one coming from the interaction between the patch and the uniform voltages, and the other due to the patch interaction. General properties of the torques are described. A convenient model of a localized patch suggested in~\cite{VA2} is used to calculate the torques explicitly in terms of elementary functions. Based on this, we analyze in detail patch interaction for one pair of patches, namely, the torque dependence on the patch parameters (width and strength) and their mutual positions. The effect of the axial torque is then studied for the experimental conditions of the STEP mission.
\end{abstract}


\maketitle

\section{Introduction\label{s1}}

The actual distribution of charges on a metal does not guarantee its surface to be an equipotential because of the impurities and microcrystal structure of the material. This phenomenon, known as patch effect (PE; for its experimantal study see paper~\cite{Dar}), is responsible for the mutual force and torque between two metallic surfaces at finite distances. The effect is the larger, the closer the surfaces, as first confirmed by the calculation of the patch effect force for two parallel conducting planes~\cite{Sp}. 

PE in a cylindrical geometry was studied in the first two parts of our paper,~\cite{VA} and \cite{VA2} (henceforth referred to as CPEI and CPEII, for "`Cylindrical Patch Effect"'), where the PE energy and force have been examined. Here we calculate the torque due to PE between two coaxial cylinders. This calculation completes the study of the electrostatic interaction for a cylindrical capacitor; our analysis largely benefits from the results found in CPEI and II.

PE is important for any precision measurement if its set-up includes conducting surfaces in a closed proximity to each other. 
 For the STEP experiment~\cite{PW, Mes, PWMes, Over}, where the differential axial motion of cylindrical test masses (TM) will be used to test the  universality of free fall to an unprecedented accuracy of about 1 part in $10^{18}$, the axial torque is of a particular interest (see section \ref{s6.}).

We determine PE torques between the two boundary surfaces of an infinitely long cylindrical capacitor. By energy conservation the rotation by an angle $\gamma$ about the direction $\hat{\gamma}$ of one of the conductors relative to the other causes an  electrostatic torque in the same direction which is given by the formula (see, for instance,~\cite{Smy})
\be
T_\gamma = -\frac{\partial W(\gamma)}{\partial \gamma}\; ,
\label{torquegen.}
\ee
where $W(\gamma)$ is the electrostatic energy as a function of the rotation angle. However, due to the specifics of cylindrical geometry, we can properly imagine a rotation only about the symmetry axis of the two infinite cylinders. Tilting, say, the inner one about any other direction leads to the intersection of the cylinders at some finite distance, i. e., to the breaking of the problem geometry. For this reason, we employ also a different method of the torque calculation. The force, $d\vec{F}$, due to the electrical field $\vec{E}$ acting on a small area $dA$ of a conductor with the charge density $\sigma$ is given by
\be
\label{forcegen.}
d\vec{F}=\sigma\vec{E}dA\;,
\ee 
(see our comment~\cite{foot1}). The resulting element of the torque about the origin at distance $\vec{r}$ from $dA$ is then
\be
d\vec{T}=\vec{r}\times d\vec{F}\;,
\label{torquegen2.}
\ee   
which expression, integrated all over the body surface, gives the general expression of the torque acting on the conductor.

The energy, the field and the surface charge density which are needed in the formulas (\ref{torquegen.}) and (\ref{torquegen2.}) are expressed through the electrostatic potential in the gap. For typical experimental conditions, such as in the STEP configuration ~\cite{Mes, Over}, the gap, $d=b-a$, is much smaller than either of the cylinder radii, $a<b$. This, first, justifies the model of infinite cylinders, especially if the patches are predominantly far from the real cylinder edges, and, second, it allows for a significant simplification of results to lowest order in $d/a$.

In the next section we solve the boundary value problem (BVP) for the potential in the gap with general voltage distributions on the cylinder surfaces. Based on this, we find the energy in section \ref{s3.}, and then the longitudinal PE torque by the formula (\ref{torquegen.}). In section \ref{s4.} we derive all the three components of the PE torque by the surface integration method using formula (\ref{torquegen2.}) (the two expressions for the axial torque agree precisely). In section \ref{s5.} a model of the localized patch potential introduced in CPEII is described, ending with closed--form expressions for the torques. The latter are then calculated and analyzed in the case when a single patch is present at each of the cylinders. In section \ref{s6.} we apply our results to the STEP experiment set--up, coming up with a clear picture of the test mass rotation motion. The details of calculations are given in two appendices.

\section{Electrostatic potential\label{s2.}}

\begin{figure}[ht]
\centering
\includegraphics[scale=0.45]{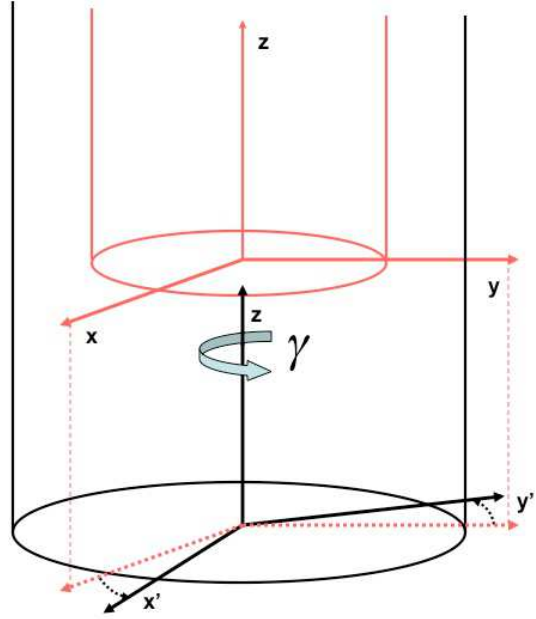}
\caption{{Geometry of the problem and coordinate systems}}
\label{fig1}
\end{figure}

We employ both Cartesian and cylindrical coordinates in two frames related to the inner and outer cylinders as shown in fig. \ref{fig1}. In the outer frame an arbitrary point is labelled by the vector radius $\vec{r}\;^{'}$, Cartesian coordinates $\{x^{\;'},y^{\;'},z\}$, or cylindrical coordinates $\{\rho,\vfi^{\;'},z\}$; in the inner frame the corresponding quantities are $\vec{r}$, $\{x,y,z\}$,  $\{\rho,\vfi,z\}$. The origins of the frames coincide so that the primed and unprimed coordinates are related simply by a rotation, by some angle $\gamma$, about the ${z}$ axis:
\be
x\;^{'} = x\cos{\gamma}+y\sin{\gamma},\quad y\;^{'} = -x\sin{\gamma}+y\cos{\gamma}, \quad z\;^{'} = z\; ,
\label{coord_rot.}
\ee
or, in cylindrical coordinates, $\rho'=\rho,\; \vfi'=\vfi -\gamma,\; z\,^{'} = z$. The surfaces of the inner and outer cylinders are thus described by the equations $\rho=a$ and $\rho=b$, respectively, and are assumed to carry arbitrary distributions of electrostatic voltage. Hence the electrostatic potential, $\Phi$, satisfying the Laplace equation in the gap between the cylinders,
\be
\Delta\Phi =0,\qquad \rho>a,\quad \rho<b,\quad 0\leq\vfi<2\pi,\quad |z|<\infty\; ,
\label{Lapl.}
\ee 
satisfies also the boundary conditions of the first kind at the cylinder surfaces:
\be
\Phi\biggl|_{\rho=a}=G(\vfi,z)\; ;\qquad\quad
\Phi\biggl|_{\rho=b}\;\;=\Vm+H(\vfi',z)=\Vm+H(\vfi-\gamma,z)\;.
\label{bc.}
\ee
Here $\Vm=const$ is the uniform potential difference, so all  voltages are counted from the uniform voltage of the inner cylinder taken as zero. The non--uniform potential distributions, i.e., the patch voltages, are described by arbitrary smooth enough functions $G(\vfi,z)$ and $H(\vfi,z)$. Same as in CPEI, II we assume these functions  squarely integrable; wherever proper, we will also assume them, as done in conditions (10), (A11) (A12) and (C5) of CPEI.

For any squarely integrable function $u(\vfi,z)$ we have its Fourier expansion with the Fourier coefficient $u_n(k)$:
\bea
u(\vfi,z)=\LL\,u_n(k)e^{i(kz+n\vfi)}\; ,\label{Four.}\quad
u_n(k)=\FF\,u(\vfi,z)e^{-i(kz+n\vfi)}\;.\;
\eea
For any two such functions $u(\vfi,z)$ and $v(\vfi,z)$ the useful Parceval identity holds,
\be
(u,\,v)\equiv\int\limits_{0}^{2\pi}\int\limits_{-\infty}^{\infty}d\vfi\, dz\,u(\vfi,z)v^*(\vfi,z)= 
\LLnof u_n(k)v^*_n(k)\; ;
\label{Parc.}
\ee
here and elsewhere the star denotes complex conjugation. 

According to the boundary condition (\ref{bc.}), we split the potential in two parts due to, respectively, the uniform boundary voltages and patches:
\be
\Phi(\vec r)\; =\; \Fiu(\vec r)\;+\;\Fip(\vec r)\; ,
\label{Fiup.}
\ee 
\be
\Fiu\biggl|_{\rho=a}=0,\qquad
\Fiu\biggl|_{\rho=b}=\Vm\; ;
\label{bcu}
\ee
\bea
\Fip\biggl|_{\rho=a}=G(\vfi,z)=\frac{1}{2\pi}\,\LLnof{G_n(k)}e^{i(kz+n\vfi)},\label{bcp}\qquad\qquad\;\;\\
\Fip\biggl|_{\rho=b}\;\;=H(\vfi-\gamma,z)=\frac{1}{2\pi}\,\LLnof{H_n(k)e^{-in\gamma}}e^{i(kz+n\vfi)}\; ;
\nonumber
\eea
Function $\Fiu$ is the classical solution for a cylindrical capacitor~\cite{Smy}; to l. o. in $d/a$ it is
\be
\Fiu(\vec{r})=\left(a/d\right)\Vm\,\ln\left(\rho/a\right)\;.
\label{Fiu.}
\ee 
Function $\Fip$ is gotten by the standard separation of variables in cylindrical coordinates [see c.f.~\cite{Leb}, Chs. 5, 6].
Its representation satisfying formally the Laplace equation is:
\be
\Fip(\vec{r})=\LL\left[A_n(k)\,I_n(k\rho)+B_n(k)\,K_n(k\rho)\right]e^{i(kz+n\vfi)}\; ,
\label{preprpr.}
\ee 
where $I_n(\xi),\;K_n(\xi)$ are the modified Bessel functions of the 1st and 2nd kind, respectively [the Macdonald's function $K_n(\xi)$  definition for the negative values of its argument is taken by the parity of $I_n(\xi)$; so, $K_n(k\rho)$ stands for $(\mbox{sign}\;k)^n K_n(|k|\rho)$].
The unknown $A_n(k),\;B_n(k)$ are found form the linear system implied by the boundary conditions (\ref{bcp}):
\bea
A_n(k)\,I_n(ka)\;+\;B_n(k)\,K_n(ka)\;=\;G_n(k)\; ;\qquad\qquad\qquad\qquad\qquad\qquad\\
A_n(k)\,I_n(kb)\;+\;B_n(k)\,K_n(kb)\;=\;H_n(k)e^{-in\gamma},\qquad n=0,\pm1,\pm2,\ldots\; .\nonumber
\label{1stbc.}
\eea
It is the same system that has been effectively solved, to lowest order in $d/a$, in CPEI [Appendix A, coefficients $A_n^0(k)$ and $B_n^0(k)$], with the exception of $e^{-\imath\,n\gamma}$ in the r.h.s. instead of $e^{\imath kz^0}$. So, by replacing $H_n(k)e^{-ikz^0}$ with $H_n(k)e^{-\imath\,n\gamma}$ in the answer (A13), CPEI, we get:
\bea
\label{AnBnsimp.}
A_n(k)=-\frac{a}{d}\left\{
K_n(kb)\left[G_n(k)-H_n(k)e^{-\imath\,n\gamma}\right]
\right\}\; ,\\
B_n(k)=\frac{a}{d}\left\{
I_n(ka)\left[G_n(k)-H_n(k)e^{-\imath\,n\gamma}\right]
\right\}\; .\qquad\nonumber
\eea
Thus the electrostatic potential (\ref{preprpr.}), to l. o. in $d/a$, is:
\be
\Fip(\vec{r})=-\frac{a}{d}\int_{-\infty}^{\infty}dk\sum_{n=-\infty}^{\infty}
\left[G_n(k)-H_n(k)e^{-\imath\,n\gamma}\right]\Omega_n(k\rho)e^{\imath\left(kz+n\vfi\right)}\;;
\label{Fip.}\quad
\ee
\be
\Omega_n(k\rho)=K_n(kb)I_n(k\rho)-I_n(ka)K_n(k\rho)\, .
\label{Omega.}
\ee
Formulas (\ref{Fiu.}), (\ref{Fip.}), and  (\ref{Omega.}) allow us to calculate both the electric energy and field.

\section{Axial Torque by the Energy Method\label{s3.}}

The uniform potential (\ref{Fiu.}) does not depend on $\gamma$, so the variation of the electrostatic energy due to the rotation comes only from the patch potential (\ref{Fip.}), same as it happens with the axial PE force (CPEII, section III). The axial torque, $T_z$, is thus given by the formula (\ref{torquegen.}) where $W(\gamma)$ is replaced with $W^p(\gamma)$ that we calculate below.

\subsection{Electrostatic Energy\label{s3.1.}}

Denote ${\cal D}_{\infty}$ the infinite domain between the two cylinders of our capacitor. The patch energy stored there, finite due to the locality of patch distributions, is: 
\bea
W^{p}=\frac{\epsilon_0}{2}\,\int_{{\cal D}_\infty}\left(\nabla \Fip\right)^2\,dV
=\nonumber\qquad\qquad\qquad\qquad\qquad\qquad\qquad\qquad\qquad\qquad\qquad\\
\frac{\epsilon_0}{2}\left\{
\int_{-\infty}^{\infty}dz\int^{2\pi}_0 b d\vfi\,H(\vfi-\gamma,z)\frac{\partial \Fip}{\partial \rho}\Biggl|_{\rho=b}-
\int_{-\infty}^{\infty}dz\int^{2\pi}_0 a d\vfi\,G(\vfi,z)\frac{\partial \Fip}{\partial \rho}\Biggl|_{\rho=a}
\right\}\;.\label{energyp}\quad
\eea
Here we used boundary conditions (\ref{bcp}) and the fact that the potential is harmonic in the domain ${\cal D}_{\infty}$, see section III in CPEI for details. The double integrals above are calculated via Fourier coefficients of the potential and its derivative in $\rho$ by the Parceval identity (\ref{Parc.}). The Fourier coefficients of the derivatives are found from the formula (\ref{Fip.}); the calculation goes the same way as in CPEI, Appendix C, and results in
\be
\label{Fipder}
\frac{\partial \Fip}{\partial \rho}\Biggl|_{\rho=a,b}=
-\frac{1}{d}\int_{-\infty}^{\infty}dk\sum_{n=-\infty}^{\infty}
\left[G_n(k)-H_n(k)e^{-\imath\,n\gamma}\right]e^{\imath\left(kz+n\vfi\right)}\;.
\ee
To l. o. in $d/a$, this expression holds at both the inner and the outer boundary. Using the Fourier coefficients, $G_n(k)$ and $H_n(k)$, of the boundary functions, we write formula (\ref{energyp}) as: 
\be
\label{Wp.}
W^{p}=\frac{\epsilon_0 a}{2d}\int_{-\infty}^{\infty}dk\sum_{n=-\infty}^{\infty}\Bigl|G_n(k)-H_n(k)e^{-\imath n\gamma}\Bigr|^2\;.
\ee
The only part of this that depends on $\gamma$, and thus contributes to the axial torque, is:
\be
\label{Wpgamma}
W^{p}(\gamma)=-\frac{\epsilon_0 a}{d}\int_{-\infty}^{\infty}dk\sum_{n=-\infty}^{\infty}\Re\Bigl[G_n(k)H^{*}_n(k)e^{\imath n\gamma}\Bigr]\;.
\ee

\subsection{Axial Torque\label{s3.2.}}

Using (\ref{Wpgamma}), we calculate the axial torque by the formula (\ref{torquegen.}):
\be
\label{Tgamma}
T_z=-\frac{\partial W^p(\gamma)}{\partial \gamma}=
-\frac{\epsilon_0 a}{d}\int_{-\infty}^{\infty}dk\sum_{n=-\infty}^{\infty}n\,\Im\Bigl[G_n(k)H^{*}_n(k)e^{\imath n\gamma}\Bigr]\;.
\ee
This representation is valid, to lowest order in $d/a$, for an arbitrary rotation $\gamma$. The torque  does not vanish only if patches are at both boundaries [$G_n,\;H_n\not\equiv0$]. A non-zero torque is generally found for $\gamma=0$, unless $G_n(k)=\lambda H_n(k)$, $\lambda$ real, i. e., the patch distributions at both cylinders are the same up to scaling. The expression of the axial torque perfectly matches that of the axial force in the symmetric configuration [CPEII, formula (21)]; the only difference is $\gamma$ instead of $z^0$, and the factor $n$ in place of $k$.    

\section{All Torques by the Surface Integration Method\label{s4.}}

\subsection{General Formulas for Electrostatic Torques\label{s4.1.}}

According to the formula (\ref{torquegen2.}), the patch effect torque on the outer cylinder is:
\bea
\label{TorqueCyl}
\vec{T}=b\int_0^{2\pi} d\vfi \int_{-\infty}^\infty dz\,\sigma\left(\vec{r}\times\vec{E}\right)\Biggl|_{\rho=b}=\qquad\qquad\qquad\qquad\qquad\qquad\qquad\qquad\qquad\\
\epsilon_0 a\int_0^{2\pi} d\vfi \int_{-\infty}^\infty dz \Biggl[\vec{r}\times\left(\frac{\partial \Phi}{\partial \rho}\hat{e}_\rho+\frac{1}{\rho} \frac{\partial \Phi}{\partial \vfi}\hat{e}_\vfi+\frac{\partial \Phi}{\partial z}\hat{e}_z\right) \frac{\partial \Phi}{\partial \rho}\Biggr]\Biggl|_{\rho=b}\left[1+O\left(\frac{d}{a}\right)\right]\;.\nonumber
\eea
Here we expressed the electrical field through the potential, and the charge density as the product of $\epsilon_0$ and the normal component of the field, by the Gauss law. We also set \break $b=a+d\approx a$, so the above holds to l. o. in $d/a$. The Cartesian components of the torque are found using the well--known cylindrical unit vectors:
\bea
\label{TorqueCar}
\vec{T}=\epsilon_0 a\int_0^{2\pi} d\vfi \int_{-\infty}^\infty dz \,\Biggl\{
\left[-\frac{z}{\rho} \frac{\partial \Phi}{\partial \vfi}\cos{\vfi}+\left(\rho\frac{\partial \Phi}{\partial z}-z\frac{\partial \Phi}{\partial \rho} \right)\sin{\vfi}\right]\hat{x}\,-\\
\left[\frac{z}{\rho} \frac{\partial \Phi}{\partial \vfi}\sin{\vfi}+\left(\rho\frac{\partial \Phi}{\partial z}-z\frac{\partial \Phi}{\partial \rho} \right)\cos{\vfi}\right]\hat{y}\,+
\left(\frac{\partial \Phi}{\partial \vfi} \right)\hat{z}\Biggr\}\frac{\partial \Phi}{\partial \rho}\Biggl|_{\rho=b}\;.\nonumber
\eea
Because of the bilinear structure of this expression, the splitting (\ref{Fiup.}) of the potential in the sum of  $\Fiu$ and $\Fip$ implies formally three contributions to the torque: one from the uniform voltages only, the other due to the interaction of patches and uniform voltages, and the third one from the patches only, just like we had it for the force in CPEII. However, uniform voltages do not give any torque, so the first contribution vanishes. Likewise, the interaction between the patches and uniform potential difference gives zero axial torque: the interaction torque is perpendicular to the symmetry axis (slanting torque):
\be
\label{IntTorquex}
T_x^{Int}=\epsilon_0 a\frac{\partial \Phi^U}{\partial \rho}\Biggl|_{\rho=b}\int_0^{2\pi} d\vfi \int_{-\infty}^\infty  dz \,
\left(-2z \frac{\partial \Phi^p}{\partial \rho} \sin{\vfi}-\frac{z}{\rho} \frac{\partial \Phi^p}{\partial \vfi}\cos{\vfi}+\rho\frac{\partial \Phi^p}{\partial z}\sin{\vfi}\right)
\Biggl|_{\rho=b}\;;
\ee
\be
\label{IntTorquey}
T_y^{Int}=\epsilon_0 a\frac{\partial \Phi^U}{\partial \rho}\Biggl|_{\rho=b}\int_0^{2\pi} d\vfi \int_{-\infty}^\infty  dz \,
\left(2z \frac{\partial \Phi^p}{\partial \rho} \cos{\vfi}-\frac{z}{\rho} \frac{\partial \Phi^p}{\partial \vfi}\sin{\vfi}-\rho\frac{\partial \Phi^p}{\partial z}\cos{\vfi}\right)
\Biggl|_{\rho=b}\;.
\ee

In contrast with that, the torque due to the patch interaction generally has all the components:
\be
\label{pTorquex}
T_x^{p}=\epsilon_0 a\int_0^{2\pi} d\vfi \int_{-\infty}^\infty  dz \Biggl\{
\frac{\partial \Phi^p}{\partial \rho}\left[-\frac{z}{\rho} \frac{\partial \Phi^p}{\partial \vfi}\cos{\vfi}+\left(\rho\frac{\partial \Phi^p}{\partial z}-z\frac{\partial \Phi^p}{\partial \rho} \right)\sin{\vfi}\right]\Biggr\}
\Biggl|_{\rho=b}\;;
\ee
\be
\label{pTorquey}
T_y^{p}=-\epsilon_0 a\int_0^{2\pi} d\vfi \int_{-\infty}^\infty  dz \Biggl\{
\frac{\partial \Phi^p}{\partial \rho}\left[\frac{z}{\rho} \frac{\partial \Phi^p}{\partial \vfi}\sin{\vfi}+\left(\rho\frac{\partial \Phi^p}{\partial z}-z\frac{\partial \Phi^p}{\partial \rho} \right)\cos{\vfi}\right]\Biggr\}
\Biggl|_{\rho=b}\;;
\ee
\be
\label{pTorquez}
T_z^{p}=\epsilon_0 a\int_0^{2\pi} d\vfi \int_{-\infty}^\infty dz \Biggl(\frac{\partial \Phi^p}{\partial \rho}\frac{\partial \Phi^p}{\partial \vfi}\Biggr)\Biggl|_{\rho=b}\;.\qquad\qquad\qquad\qquad\qquad\qquad\qquad\quad\;
\ee
Formulas (\ref{IntTorquex})---(\ref{pTorquez}) provide the general representation for the torque based on the surface integration method. Below we use them, along with the expressions (\ref{Fiu.}) and (\ref{Fip.}) of the potential in the gap, to find the PE torque on the cylinder.

\subsection{Axial Torque by the Surface Integration Method\label{s4.2.}}

We compute the axial component (\ref{pTorquez}) of the torque employing the Parceval identity (\ref{Parc.}):
\bea
T_z^{p}=\epsilon_0 a \int_{-\infty}^\infty dk \sum_{n=-\infty}^\infty \left[-\frac{1}{2d}\left(G_n^{*}(k)-H_n^{*}(k)e^{in\gamma}\right)\,in\,H_n(k)e^{-in\gamma}+\right.\qquad\qquad\qquad\qquad\\
\left.
\frac{1}{2d}\left(G_n(k)-H_n(k)e^{-in\gamma}\right)\,in\,H_n^{*}(k)e^{in\gamma}\right]=\;\nonumber
\label{Tz}
-\frac{\epsilon_0 a}{d} \int_{-\infty}^\infty dk \sum_{n=-\infty}^\infty \Im\left[nG_n(k)H_n^{*}(k)e^{in\gamma}\right]\;. 
\eea
We used Fourier coefficients of the two derivatives of the potential, the radial one (\ref{Fipder}), and the angular one obtained by differentiating the second of boundary condition (\ref{bcp}):
\be
\label{dphidphi}
\frac{\partial \Phi^p}{\partial \vfi}\Biggl|_{\rho=b}=\frac{1}{2\pi}\,\LLnof e^{i(kz+n\vfi)}\left[in\,H_n(k)e^{-in\gamma}\right]\;.
\ee
The  axial torque (\ref{Tz}) and (\ref{Tgamma}) by the surface integration and energy method,  respectively, is exactly the same; this is an important cross--check of our calculations.

\subsection{Slanting Torques by the Surface Integration Method\label{s4.3.}}

\subsubsection{Uniform and patch potential interaction\label{s4.3.1.}}

We start with calculating the torque due to the interaction of uniform and patch potentials. We first substitute the uniform field, ${\Vm}/{d}$, in the formula (\ref{IntTorquex}):
\[
T_x^{Int}=\frac{\epsilon_0a}{d}\Vm \int_0^{2\pi} d\vfi \int_{-\infty}^\infty  dz \,
\left(-2z \frac{\partial \Phi^p}{\partial \rho} \sin{\vfi}-\frac{z}{\rho} \frac{\partial \Phi^p}{\partial \vfi}\cos{\vfi}+\rho\frac{\partial \Phi^p}{\partial z}\sin{\vfi}\right)
\Biggl|_{\rho=b}\; ,
\]
to get the $x$ component. The last term vanishes after integrating in $z$, so:
\[
T_x^{Int}=\frac{\epsilon_0a}{d}\Vm \int_0^{2\pi} d\vfi \int_{-\infty}^\infty  dz \,
\left(-2z \frac{\partial \Phi^p}{\partial \rho} \sin{\vfi}-\frac{z}{\rho} \frac{\partial \Phi^p}{\partial \vfi}\cos{\vfi}\right)
\Biggl|_{\rho=b}\;.
\]
By the definition (\ref{Four.}) of the Fourier transform, this double integral is equal to $2\pi$ times the Fourier coefficient of the integrand at $n=k=0$. The needed Fourier coefficients are determined in Appendix~A, formulas (\ref{dphidrhophisin}) and (\ref{dphidphizcos}), so:
\be
\label{TxInt}
T_x^{Int}=-4\pi\frac{\epsilon_0a}{d^2}\Vm \Re \Biggl[
\frac{\partial}{\partial k}\left(G_{1}(k)-H_{1}(k)e^{-i\gamma}\right)
\Biggr]
\Biggl|_{k=0}\left[1+O\left(\frac{d}{a}\right)\right]\;.
\ee
This final compact formula for the torque has been obtained by employing the property 
$G_{-n}(-k)=G_{n}^{*}(k),\;H_{-n}(-k)=H_{n}^{*}(k)$ of Fourier coefficients of real functions. The estimate of the remainder in the expression (\ref{TxInt}) holds for patch distributions $G(\vfi,z)$ and $H(\vfi,z)$ satisfying conditions (A11), (A12) and (C5), CPEI, and also such that the products $zG(\vfi,z)$, and $zH(\vfi,z)$ are squarely integrable, see formula (\ref{addcond1}) in Appendix A. The validity of these conditions is assumed everywhere below, including the final expressions of all the torques.

A similar calculation for the $y$ component starts from the formula (\ref{IntTorquey}) and uses the Fourier coefficients (\ref{dphidrhophicos}) and (\ref{dphidphizsin}). It results, to lowest order in $d/a$, in:
\be
\label{TyInt}
T_y^{Int}=4\pi\frac{\epsilon_0a}{d^2}\Vm \Im \Biggl[
\frac{\partial}{\partial k}\left(G_{1}(k)-H_{1}(k)e^{-i\gamma}\right)\Biggr]\Biggl|_{k=0}\; ,
\ee
with the same remainder as in formula (\ref{TxInt}).

\subsubsection{Patch potentials interaction\label{s4.3.2.}}

Now we go for the expressions of the torque caused by the interaction between patches. The $x$ component of this torque, $T^p_x$, given by the formula (\ref{pTorquex}), with the help of the Parceval identity becomes [the proper Fourier coefficients are found in (\ref{Fipder}), (\ref{dphidrhophisin}), (\ref{dphidphizcos}), and (\ref{dphidzphisin})]:
\bea
\label{pTorquexEq}
T_x^{p}=-\frac{\epsilon_0 a}{2d^2} \int_{-\infty}^\infty  dk \sum _{n=-\infty}^\infty
\Biggl[G_n^{*}(k)-H_n^{*}(k)e^{\imath\,n\gamma}\Biggr]\times\qquad\qquad\qquad\qquad\qquad\qquad\qquad\nonumber\\
\frac{\partial }{\partial k}\Biggl[\left(G_{n-1}(k)-H_{n-1}(k)e^{-i(n-1)\gamma}\right)-
\left(G_{n+1}(k)-H_{n+1}(k)e^{-i(n+1)\gamma}\right)\Biggr]\left[1+O\left(\frac{d}{a}\right)\right]\;.\nonumber
\eea
This formula can be simplified further: integrating by part in the first of the two products and shifting there the index $n$ by one, $n'=n-1$, leads to the final more compact representation:
\be
\label{Txp}
T_x^{p}=\frac{\epsilon_0 a}{d^2} \int_{-\infty}^\infty  dk \sum _{n=-\infty}^\infty\Re\Biggl\{\Biggl[G_n^{*}(k)-H_n^{*}(k)e^{\imath\,n\gamma}\Biggr]\frac{\partial}{\partial k}\Biggl[G_{n+1}(k)-H_{n+1}(k)e^{-\imath\,(n+1)\gamma}\Biggr]\Biggr\}\;. 
\ee 

The other component, $T^p_y$, can be determined in a similar way starting with the expression (\ref{pTorquey}) and combining it with the formulas (\ref{Fipder}), (\ref{dphidrhophicos}), (\ref{dphidphizsin}), and (\ref{dphidzphicos}). The result is: 
\be
\label{Typ}
T_y^{p}=-\frac{\epsilon_0 a}{d^2} \int_{-\infty}^\infty  dk \sum _{n=-\infty}^\infty\Im\Biggl\{\Biggl[G_n^{*}(k)-H_n^{*}(k)e^{\imath\,n\gamma}\Biggr]\frac{\partial}{\partial k}\Biggl[G_{n+1}(k)-H_{n+1}(k)e^{-\imath\,(n+1)\gamma}\Biggr]\Biggr\}\; , 
\ee
entirely similar to (\ref{Txp}). The general analysis of PE torques is completed. 

\subsection{General properties of the PE torque \label{s4.4.}}

Looking at the results of the calculation of the PE torques, one can come up with a few general conclusions regarding their properties, such as:
\vskip1mm

1. Uniformly charged cylinders do not give rise to any torque.
\vskip1mm

2. The axial torque is inversely proportional to the gap width, the transverse components are inversely proportional to its square.
\vskip1mm

3. Patches need to be present on both the cylinders to generate an axial torque.
\vskip1mm

4. A non-zero axial torque is generally found when the cylinders are not rotated against each other ($\gamma=0$), unless the patch voltage distributions on both cylinders are the same up to a factor.
\vskip1mm

5. Just one patch is enough to generate a slanting torque.
\vskip1mm

6. A non-zero slanting torque is generally found when $\gamma=0$ unless the patch voltage distributions on both cylinders are the same.
\vskip1mm

7. The interaction between patches and uniform potentials involves only the first harmonics of the azimuthal angle of the patch distribution.
\vskip1mm

So, the general formulas that we obtained enable one to make some significant conclusions about PE torques agreeing with the physical insights into their origin.

\section{Single Patch at Each of the Electrodes\label{s5.}: a Picture of Patch Interaction}

To study the features of PE torques we analyze them in the case when only one localized patch is found at each of the cylinders, as it was done with PE force in CPEII. To make the analysis results transparent, one needs to have the torques in a simple enough closed form, which requires some special choice of the generic patch model, a rather delicate task. The localized potential distribution that satisfies this very well has been suggested and developed in CPEII, section IV. We repeat basic facts about this model and use it to calculate the torques and examine the patch interaction.

\subsection{The Patch Model\label{s5.1.}}

The suggested model of the patch potential is:
\be
\cV(\vfi-\vfi_*, z-z_*)\equiv\cV(\vfi-\vfi_*,\lambda, z-z_*,\Delta z)=V_*\,f(z-z_*)\,u(\vfi-\vfi_*)\; ,
\label{Vpatchgen.}
\ee
where
\be
f(z)=\exp{\left[-\left(\frac{z}{\sqrt{2}\,\Delta z}\right)^2\right]}\; ,\qquad u(\vfi)=u(\vfi,\lambda)=\frac{\left(1-\lambda\right)^2}{2}\,\frac{1+\cos\vfi}{1-2\lambda\cos\vfi+\lambda^2}\;;
\label{fzufi}
\ee
[note $\Bigl|f(z)\Bigr|\leq1$, $\Bigl|u(\vfi)\Bigr|\leq1$, and $f(0)=u(0)=1$].
Here $V_*=\cV(0,0)$ is the maximum magnitude of the potential (positive or negative); the center of the patch is at $\vfi=\vfi_*,\;z=z_*$, $\Delta z$ denotes the axial size of the patch, and $\lambda $ controls its angular size. Indeed, $\lambda$ is related to $\Delta \vfi$, the angular half--width of the patch [defined as the angle for which $u$ is equal to its mean value, $u(\Delta\vfi)=u_{av}$], by the equalities:
\be
\cos\Delta\vfi={\lambda}\;,\qquad \Delta \vfi=\arccos{\lambda}\; .
\label{Deltafi.}
\ee
Fourier coefficients of the patch model (\ref{Vpatchgen.}), (\ref{fzufi}) are:
\be
\label{VpatchFour.}
\cV_n(k)\equiv\cV_n(k,\lambda,\Delta z)=V_*\tf(k)e^{-ikz_*}\,u_n e^{-in\vfi_*}\;,\qquad\tf(k)=\Delta z\exp{\left[-\left(\frac{k\Delta z}{\sqrt{2}}\right)^2\right]}\;;
\ee
\be
u_n\equiv u_n(\lambda)=\sqrt{2\pi}\,\frac{1-\lambda^2}{4\lambda}\,\lambda^{\left|n\right|},\quad n\neq 0;
\qquad\qquad u_0\equiv u_0(\lambda)=\sqrt{2\pi}\,\frac{1-\lambda}{2}\;.
\label{uonn.}
\ee
For a single patch at each of the two cylinders, the boundary functions $G(\vfi,z)$ and $H(\vfi-~\gamma,z)$ are given by the formula (\ref{Vpatchgen.}) as:
\be
\label{Onebcp.}
G(\vfi,z)=\cV(\vfi-\vfi_1,\lambda_1, z-z_1,\Delta z_1)\;;\qquad
H(\vfi-\gamma,z)=\cV(\vfi-\vfi_2-\gamma,\lambda_2, z-z_2,\Delta z_2)\;.
\ee
The torques corresponding to these distributions are calculated in Appendix B in a closed form. Here we study the patch interaction for a particular case when the sizes of the patches are identical, $\Delta z_1=\Delta z_2=\Delta z$, $\Delta \vfi_1=\Delta \vfi_2=\Delta \vfi$, and $V_1=\pm V_2=V_0$. 

\subsection{Axial Torque\label{s5.2.}} 

\begin{figure}[ht]
\centering
\includegraphics[scale=0.45]{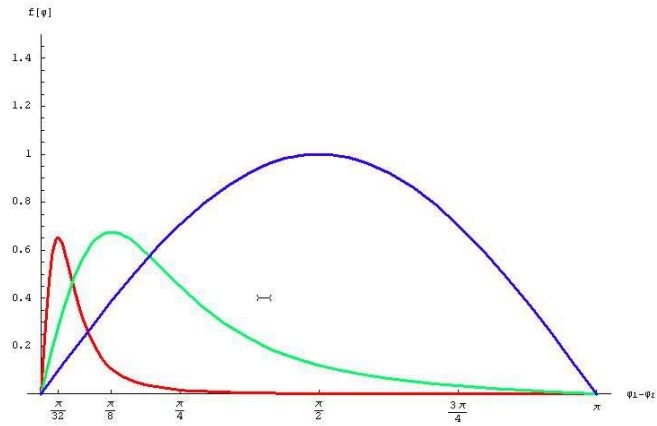}
\caption{{Axial torque vs. the angular distance between the patches for $\Delta \vfi=\pi/8\,,\pi/4\,,\pi/2$}}
\label{fig2}
\end{figure}

As shown in our general analysis, sections III and IV, the axial torque occurs only when both cylinders carry non-uniform voltages. For the case of two patches of equal sizes and magnitude, its expression is found by the formula (\ref{OneTz}) [recall that we have set $\gamma=0$]:
\be
\label{sameOneTz}
{T_z^p}=\mp\frac{\sqrt{\pi^{3}}}{4}\frac{\epsilon_0 a}{d}V_0^2\Delta z\sin^6
\left(\Delta \vfi\right)\,\mu\,\sin\left(\vfi_1-\vfi_2\right)\exp{[{-\tilde{z}^2}]}\;,
\ee 
\[
\mu\equiv\mu(\lambda,\vfi_1-\vfi_2)\equiv\frac{1+\lambda^2}
{\left[
1-2\lambda^2\cos{\left(\vfi_1-\vfi_2\right)}+\lambda^4
\right]^2}
\;;\qquad\qquad\tilde{z}\equiv\frac{z_1-z_2}{2\Delta z}\;.
\]
The signs $\mp$ in formula (\ref{sameOneTz}) stand for patches of equal or opposite potential, respectively. 
This torque is proportional to the inverse of the relative gap, $d/a$. Its dependence on the axial patch distance is driven by the Gaussian exponent: the axial torque monotonically and rapidly decreases toward zero with increasing values of $\Bigl|z_1-z_2\Bigr|$. The dependence on the angular patch distance, $0\leq\vfi_1-\vfi_2\leq\pi$ is shown in fig.\ref{fig2} for various values of $\Delta \vfi$: the torque goes down when the patch angular width gets smaller.

The axial and azimuthal sizes of the patch play a very distinctive role in the expression (\ref{sameOneTz}). As $\Delta z$ grows (and the patch thus becomes strip--like, for fixed $\Delta\vfi$), the torque magnitude increases and goes linearly to infinity in $\Delta z\to \infty$. On the other hand, the influence of the azimuthal patch width is represented predominantly by the sixth power of the sine of $\Delta \vfi$: the torque vanishes for belt--like patches, and goes to zero as $(\Delta \vfi)^6$ for $\Delta \vfi\to 0$.

\subsection{Slanting Torques\label{s5.3.}}

\subsubsection{Uniform and patch potential interaction\label{s5.3.1.}}

In our case of identical patches expressions (\ref{oneIntT}) for the interaction torque simplify to
\bea
\label{sameoneIntT}
{T_x^{Int}}=-\sqrt{2\pi^{3}}\,\frac{\epsilon_0 a}{d^{\,2}}V_0 \Vm \Delta z\sin^2(\Delta\vfi) \Bigl(z_1\sin{\vfi_1}\mp z_2\sin{\vfi_2}\Bigr)\;;\\
T_y^{Int}=\sqrt{2\pi^{3}}\,\frac{\epsilon_0 a}{d^{\,2}}V_0 \Vm \,\Delta z\sin^2(\Delta\vfi) \Bigl(z_1\cos{\vfi_1}\mp z_2\cos{\vfi_2}\Bigr)\;.\nonumber
\eea
Recall that $\Vm$ is the difference between the uniform potentials at the boundaries, see formula (\ref{bcu}).
The minus or plus sign above is taken when the patches have the same or the opposite voltages, respectively. 
Expressions (\ref{sameoneIntT}) show that this torque is a superposition of two contributions each coming from a single patch interacting with the uniform voltage $\Vm$.  Each of these torques can be expressed as the product of a force acting at the center of the patch and the respective arm, $z_1$ or $z_2$. It is interesting that these forces are equal to the zero order interaction forces obtained in CPEII, section VA. For $0\leq\Delta \vfi\leq\pi/2$ the torque grows with the azimuthal width of the patch; then, for bigger angular sizes, it decreases and goes to zero as the patch becomes annular. Moreover, for $\Delta \vfi\to 0$, it goes to zero as $(\Delta \vfi)^2$. The torque is again proportional to $\Delta z$. However, as the patch becomes strip--like, the torque goes to zero, by symmetry. In fact, the arm of the force is zero in this latter case.

\subsubsection{Patch--patch interaction \label{s5.3.2.}}

\begin{figure}[ht]
\centering
\includegraphics[scale=0.45]{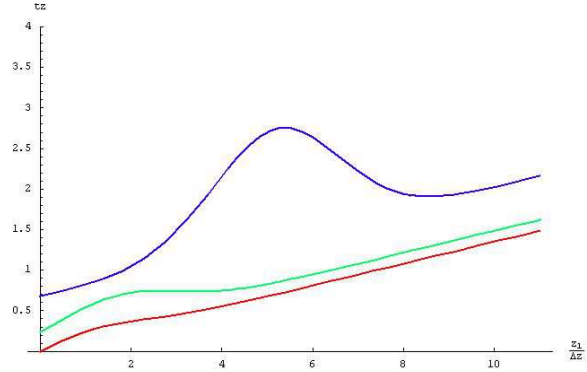}
\caption{{Slanting torque vs. the axial distance between the patches for $z_2=0\,,\Delta z\,,5\Delta z$.}}
\label{fig3}
\end{figure}

The slanting patch torque expressions (\ref{Txpr}) and (\ref{Typr}) become
\bea
\label{sameTpr}
T_x^p=-2\sqrt{\pi^{3}}\, \frac{\epsilon_0 a}{d^{\,2}}V_0^2\Delta z\sin^2{(\Delta \vfi)}\times\qquad\qquad\qquad\qquad\qquad\qquad\qquad\\
\Biggl[
{\cal N}\left(z_1\sin{\vfi_1}+z_2\sin{\vfi_2}\right)\mp \left(z_1+z_2\right){\cal M}_1\left(\sin\vfi_1+\sin\vfi_2\right)
e^{-\tilde{z}^2}
\Biggr]\nonumber\; ;\\
T_y^p=2\sqrt{\pi^{3}}\, \frac{\epsilon_0 a}{d^{\,2}}V_0^2\Delta z\sin^2{(\Delta \vfi)}\times\qquad\qquad\qquad\qquad\qquad\qquad\qquad\quad\nonumber\\
\Biggl[
{\cal N}\left(z_1\cos{\vfi_1}+z_2\cos{\vfi_2}\right)\mp \left(z_1+z_2\right){\cal M}_1\left(\cos\vfi_1+\cos\vfi_2\right)
e^{-\tilde{z}^2}
\Biggr]\nonumber\; ;
\eea
\[\tilde{z}=\frac{z_1-z_2}{2\Delta z}\;;\qquad
{\cal N}=\frac{2-\lambda}{8}\;;\quad
{\cal M}_1=\frac{1-\lambda}{8}\Bigl[
1-\frac{\lambda}{2}(1+\lambda)\frac{1+\lambda^2-2\cos{(\vfi_1-\vfi_2)}}{1-2\lambda^2\cos{(\vfi_1-\vfi_2)}+\lambda^4}
\Bigr]\
\;,
\] 
in our case of identical patches. Above we have used a slightly different notation for ${\cal M}_1$ and ${\cal N}$ as compared to Appendix B, formulas (\ref{MN}). The first term in the square brackets carries the contributions of each single patch independent of their signs. The second term represents interaction between the patches. It decreases as the distance between them grows, due to the presence of the coefficient ${\cal M}_1$ and the Gaussian exponent. The decay is faster, the smaller the widths $\Delta z$ and $\Delta \vfi$ are. The plus or minus sign of this term is taken depending on whether the patch potentials have the same or opposite signs. Similarly to the previous case, the patch torque can be expressed as the sum of the products of a force acting on each patch and the arm $z_1$ or $z_2$, plus a mutual force between the patches times the arm $z_1+z_2$ [see fig.\ref{fig3} giving torque versus $ z_1$ for different values of $z_2$]. The slanting patch torque is essentially proportional to the square sine of $\Delta \vfi$, and to $\Delta z$. However, by symmetry, as $\Delta z\to\infty$, the torque vanishes, just like the interaction torque above.

\section{Axial Patch Effect Torques for STEP\label{s6.}}

\begin{figure}[ht]
\centering
\includegraphics[scale=0.45]{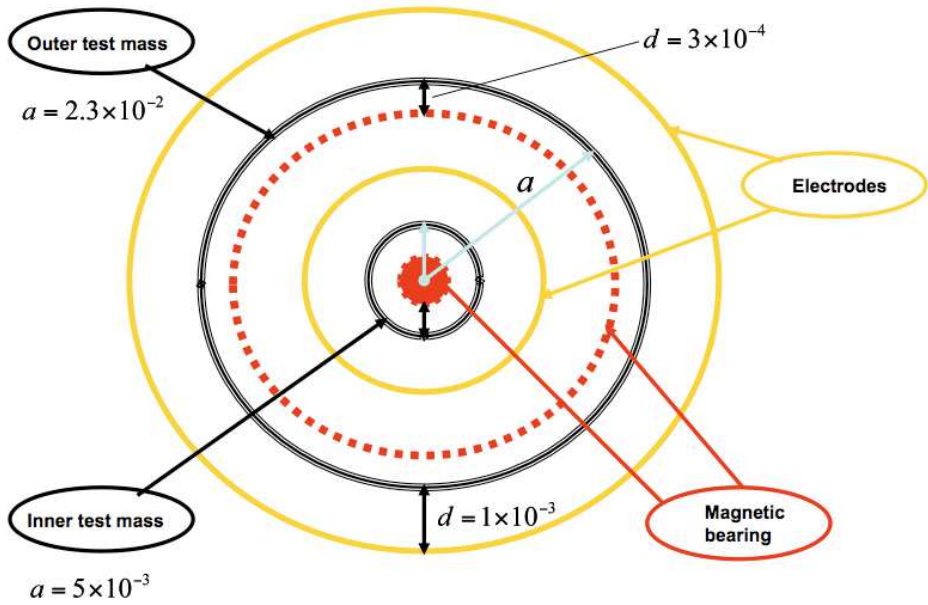}
\caption{{Cross section of the STEP differential accelerometer (all dimensions in meters)}}
\label{fig4}
\end{figure}

In this section we use the obtained results to evaluate the effect of non--uniform potentials on the performance of the instrument that will be used in STEP. The pertinent information about this experiment is found in CPEII, section VIA; for more details, see~\cite{PW, Mes, Over}. Here we recall the basic design of its core system, the differential accelerometers (DACs). As shown in fig.\ref{fig4}, each DAC consists of two test masses (TMs) shaped as coaxial cylindrical shells, and of a system of electrodes and magnetic bearings. It essentially constrains  the TMs in four degrees of freedom, with the translation along their symmetry axis and the spin about it left free. The science signal is read by a magnetic SQUID readout system from the differential axial motion of the TMs. 

In CPEII we analyzed the patch effect on the axial translation of the TMs. Here we concentrate on the axial PE torque causing the spin motion, i. e., relative rotation of the cylinders about their common axis. The main reason to examine it is that this motion makes PE forces change with the time. Just like in CPEII, we consider a TM and its magnetic bearing as a reference case of our pair of cylinders, since the gap between them is at least 3 times smaller - and the axial torque thus 3 times larger - than the gap between the TM and the electrodes. As was assumed in section \ref{s5.}, we consider each of the cylinders having just one patch, both patches of identical sizes and magnitudes. Even in this case the equation of motion proves to be a nonlinear one with the periodic potential such that two equilibrium points exist within a single period, one stable and one unstable. We describe the general picture of motion pointing out the regimes expected as typical in the real situation of the STEP experiment, namely, the regimes of oscillations. We give the expressions for the frequency of small oscillations near stable equilibria and the runaway time from the unstable ones, and then estimate these quantities under the STEP conditions. An estimate of the torque is also provided. Finally, we show that spin motion for any patch voltage distribution will be qualitatively the same as in the case of two patches only.

In compliance with CPEII, we use the following parameters: patch voltage $V_0=10mV$, TM radius $a=2.3 cm$, TM to magnetic bearing gap $d=0.3 mm$, TM length $2L=0.14 cm$. The moments of inertia for some flight--like TMs vary from $2.34 \times 10^{-5} Kg\, m^2$ to $1.40 \times 10^{-3} Kg\, m^2$ for different test masses, but a single TM is fabricated in such a way that the principal moments of inertia are all equal to a high accuracy~\cite{L}, ${\cal I}_x={\cal I}_y={\cal I}_z={\cal I}$. We here need, in fact, only ${\cal I}={\cal I}_z$, and we use the smallest value, ${\cal I}=2.34 \times 10^{-5} Kg\, m^2$.

\subsection{Spin Motion in the Case of Two Patches \label{s6.1.}}

We take the torque expression (\ref{sameOneTz}), and write the motion equation,
\be
\label{eqmot}
{\In}\,\ddot{\gamma}={T_z}=\mp\,T_*(\tilde z)\,\Delta z\sin^6\left(\Delta \vfi\right)\,\mu(\gamma,\lambda)\,\sin{\gamma}\; ,
\ee 
for the rotation of the outer cylinder about the symmetry axis; here the dimesionless function
\be\label{mugamlam}
\mu=\mu(\gamma,\lambda)=\frac{1+\lambda^2}
{\left(
1-2\lambda^2\cos{\gamma}+\lambda^4
\right)^2}>0\; ,
\ee
and the characteristic value of the torque is 
\be\label{T*}
T_*(\tilde z)=\frac{\pi^{3/2}}{4}\,\epsilon_0 V_0^2\,\frac{ a}{d}\,e^{-\tilde{z}^2},\qquad
\tilde{z}=\frac{z_1-z_2}{2\Delta z}\;.
\ee
Without any loss of generality, we count here the rotation angle $\gamma$ from the position where two patches are right one against the other, $\vfi_1-\vfi_2=0$. The torque has the minus (plus) sign when the patch voltages have the same (opposite) sign. 
Equation (\ref{eqmot}) strongly resembles the classical motion equation of the pendulum,  with just one additional coefficient $\mu(\gamma,\lambda)$ being a strictly positive non-singular function of $\gamma$. Extending this similarity, we note that, by the expression (\ref{Tgamma}) for the axial torque through the patch energy $W^p(\gamma)$, equation (\ref{eqmot}) has the potential $W^p(\gamma)/\In$: it can be equivalently written as
\be
\label{eqmotpot}
{\In}\,\ddot{\gamma}=-\frac{\partial W^p(\gamma)}{\partial\gamma}\; .
\ee 
Since the potential is periodic in $\gamma$, the complete qualitative picture of motion is well known (see, for instance,~\cite{SUZ}, Ch.1). It follows from the energy integral of the equation (\ref{eqmotpot}),
\be
\label{1stint}
\frac{{\In}\,\dot{\gamma}^2}{2}+{W^p(\gamma)}=E_0\; ,
\ee
where $E_0$ is the total energy determined by the initial conditions:
\[
E_0=\frac{{\In}\,\dot{\gamma_0}^2}{2}+{W^p(\gamma_0)},\qquad \gamma_0=\gamma(t_0),\quad \dot{\gamma_0}=\dot{\gamma}(t_0)\; .
\]
Potential energy $W^p(\gamma)$ is bounded, with the bounds denoted as 
\be\label{Wpm}
-\infty<W_-=\min\limits_{0\leq\gamma<2\pi}W^p(\gamma)\,<\, W_+=\max\limits_{0\leq\gamma<2\pi}W^p(\gamma)<\infty \; .
\ee
The minimum potential energy $W_-=W^p(\gamma_-)$ corresponds, of course, to a stable equilibrium \break $\gamma(t)\equiv\gamma_-=\mbox{const}$, 
while the maximum one, $W_+=W^p(\gamma_+)$, is achieved at an unstable equilibrium point, $\gamma(t)\equiv\gamma_+=\mbox{const}$. This is enough to qualitatively describe the motion.

Indeed, from the energy conservation (\ref{1stint}) it is clear that $E_0\geq W_-$. If $E_0= W_-$, then the system stays at the stable equilibrium, ${\gamma}(t)\equiv\gamma_0=\gamma_-,\;\dot{\gamma}(t)\equiv0,\; t\geq t_0$. If $W_-<E_0< W_+$, then the system can never reach the peak of the potential, the rotation angle is bounded at all times, $\gamma_-\leq\gamma(t)<\gamma_{max}<\gamma_+$, the motion is finite, which means that the cylinder oscillates about the stable equilibrium. If, next, $E_0> W_+$, then the system always remains above all the potential wells, the motion is infinite, the cylinder rotates indefinitely and non-uniformly in one direction depending on the sign of the initial velocity. What remains is the exceptional case $E_0=W_+$, when, in purely mathematical view, the system stays at the equilibrium $\gamma_+$; however, this rest point is unstable, so in reality any small perturbation in this or that direction leads, again, either to the oscillational, or to the rotational motion.

To make all this even more particular in our case of just two patches, we give the patch energy explicitly, as easily found either by the general expression (\ref{Wpgamma}) or, up to an insignificant constant, by the direct integration of the torque in the r.h.s. of the equation (\ref{eqmot}):
\be
W^p(\gamma)=\mp\,T_*(\tilde z)\Delta z\sin^2\left(\frac{\Delta \vfi}{2}\right)\,w(\gamma)+\mbox{const},\quad
w(\gamma)=\frac{\left(1+\lambda\right)^2}{2\lambda^2}\frac{1}{1-2\lambda^2\cos{\gamma}+\lambda^4}\; .
\label{2patchen}
\ee
There is no essential difference between the two cases with the opposite signs, so we discuss only the case of the minus sign below.

\begin{figure}[ht]
\centering
\includegraphics[scale=0.45]{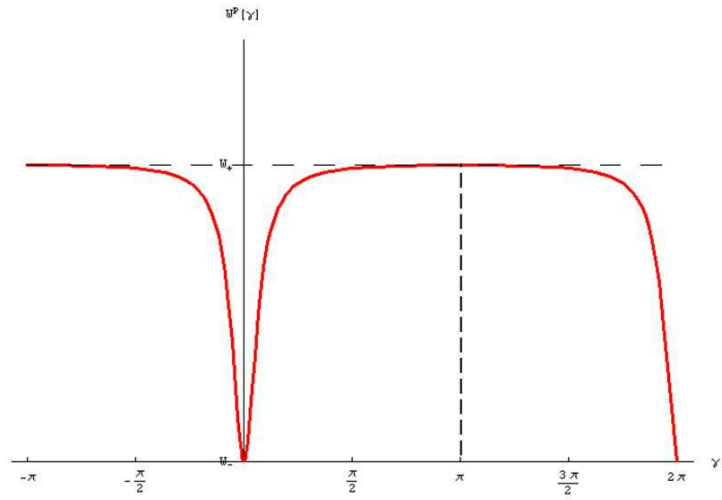}
\caption{{Periodic potential}}
\label{fig5}
\end{figure}

As seen yet from the equation (\ref{eqmot}), in this case we have just two equilibria at each period of the potential, the stable one, $\gamma_-=0\;(\mbox{mod}\, 2\pi)$, and the unstable $\gamma_+=\pi\;(\mbox{mod}\, 2\pi)$. This is also clear from the plot of the patch energy given in fig.\ref{fig5}; the horizontal line through the peaks shows the critical energy, $W_+$, that separates the finite motions (oscillations) from the infinite ones (rotations). The period of oscillations is the larger, the higher energy $E_0$ is, i.e., the larger the oscillation amplitude (the limit of small oscillations, when the period is independent of the amplitude, is described in the next section).

\begin{figure}[ht]
\centering
\includegraphics[scale=0.45]{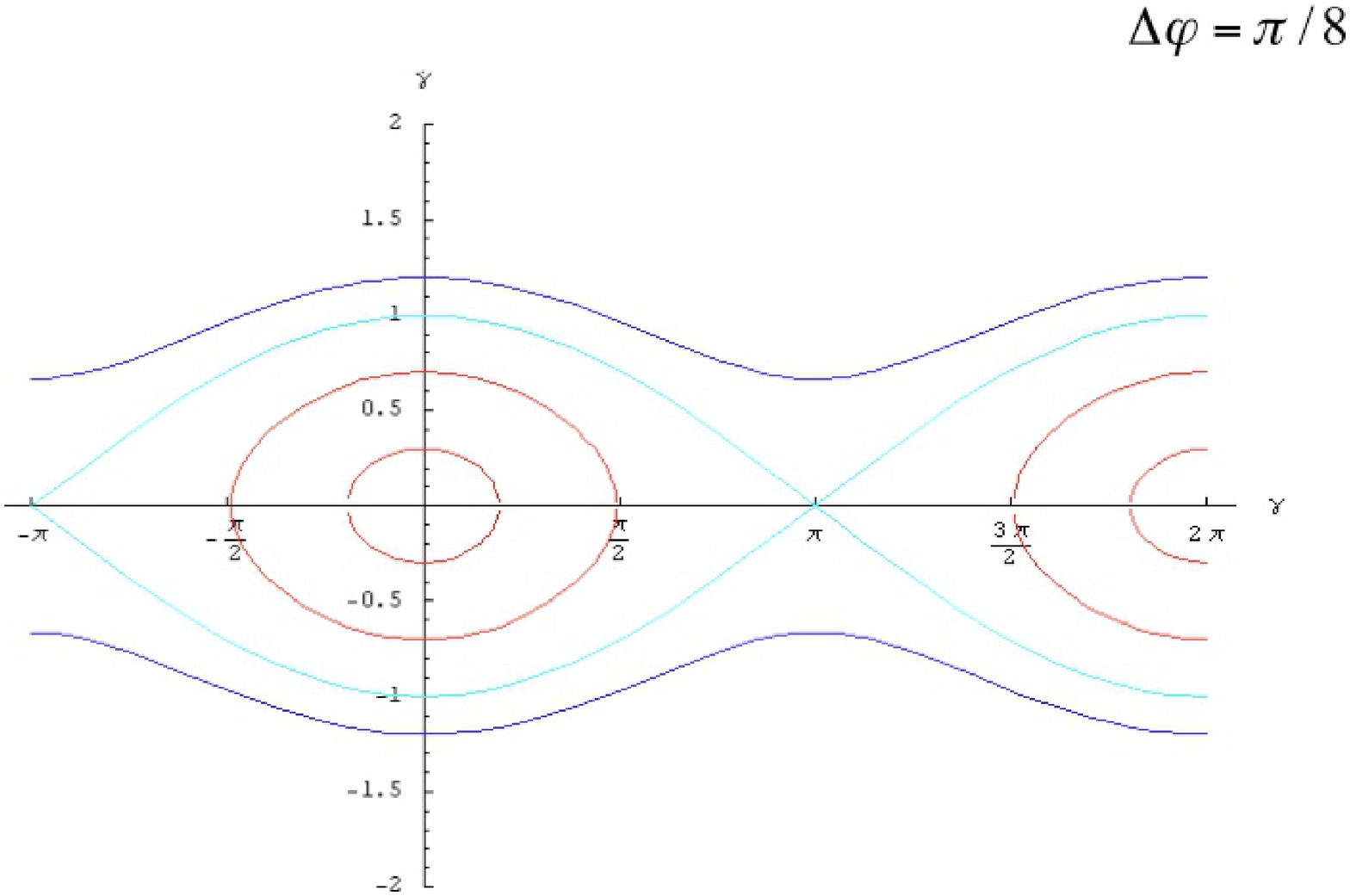}
\caption{{Phase plane of spin motion}}
\label{fig6}
\end{figure}

The energy integral (\ref{1stint}) allows also for the representation of motion in the phase plane $\gamma,\;\dot{\gamma}$. The corresponding plot is given in fig.\ref{fig6}; closed orbits in it correspond to oscillatory (finite) motions (the size of these ovals grows with $E_0$). They are separated from the infinite trajectories (rotations) by the so called heteroclinic curves, which go from one unstable equilibrium to another nearest to it. The total time of motion along these separatrices from one rest point to the other is infinite.

Note that under the STEP conditions one expects only oscillatory spinning of the test mass, rather than its rotation. The reason is that the TM will be caged (fixed) during the satellite launch, and then rather accurately released, practically with no initial velocity, which leads to pure oscillations.

\subsection{Estimates for Small TM Motions Near Its Equilibria in the Case of Two Patches \label{s6.2.}}

Here we describe small motions of the cylinder near its rest points. Accordingly, we linearize the equation (\ref{eqmot}) by setting
\[
\gamma(t)=\gamma_\mp+\delta \gamma(t),\qquad |\delta\gamma(t)|\ll 1\; ,
\]
which results in
\be
\label{eqmotlin}
\ddot{\delta \gamma}=\mp\,\omega^2_{\mp}\,\delta \gamma,\qquad 
\omega^2_{\mp}= \frac{T_*(\tilde{z})}{\In }\,\Delta z\sin^6\left(\Delta \vfi\right)
\frac{\left(1+\lambda^2\right)}
{\left(1\mp\lambda^2\right)^4}\;.
\ee
The cylinder thus oscillates about the stable equilibirium position $\gamma_-=0$ with the frequency
\be
\label{oscfreq}
f^{PE}=\frac{\omega_-}{2\pi}=
\frac{\pi^{-1/4}V_0}{4}\sqrt{\frac{\epsilon_0 a}{\In d}}\sqrt{\Delta z}\frac{\sqrt{1+\cos^2{\Delta \vfi}}}{\sin{\Delta \vfi}}
e^{-0.5{\tilde z}^{\,2}}\leq 10^{-5}\sqrt{\Delta z}\frac{\sqrt{1+\cos^2{\Delta \vfi}}}{\sin{\Delta \vfi}}\,Hz\; .
\ee
Accordingly, it rotates exponentially away from the unstable position $\gamma_+=\pi$ with the characteristic time
\be
\label{runaway}
\tau^{PE}=\frac{1}{\omega_+}=
\frac{2}{\pi^{3/4}V_0}\sqrt{\frac{\In d}{\epsilon_0 a}}\frac{1}{\sqrt{\Delta z}}\frac{\left(1+\cos^2{\Delta \vfi}\right)^{3/2}}{\sin^3{\Delta \vfi}}
e^{0.5{\tilde z}^{\,2}}\geq 1.5\times10^{4}\frac{\left(1+\cos^2{\Delta \vfi}\right)^{3/2}}{\sqrt{\Delta z}\,\sin^3{\Delta \vfi}}\,s\;.
\ee
Numerical estimates (\ref{oscfreq}) and (\ref{runaway}) hold for the above set of STEP parameters, with $\Delta z$ in meters. To get the feeling of what the numbers are in reality, let us consider some examples. For instance, the longitudinal patch size cannot be larger than the size of the TM, $\Delta z\leq L$; in the case of the maximum size, the upper bound for the frequency of small oscillations becomes:
\[
f^{PE}\leq 2.6\times10^{-6}\frac{\sqrt{1+\cos^2{\Delta \vfi}}}{\sin{\Delta \vfi}}\,Hz\; .
\]
This value remains below the STEP signal frequency range 
\[
1.74\times 10^{-4}Hz<f<5.2\times 10^{-4}Hz\ ;
\]
(see CPEII, section VIA) for the patches as small in the azimuthal direction as $\Delta \vfi\sim 1\,deg$, or larger. The frequency tends to infinity when the angular size tends to zero, because the potential well becomes infinitely deep.
The smallest runaway time in the case $\Delta z=L$ is
\[
\tau^{PE}\geq 5.6\times10^{4}\frac{\left(1+\cos^2{\Delta \vfi}\right)^{3/2}}{\sin^3{\Delta \vfi}}\,s\; ,
\]
which is much larger than the duration of a single STEP science session, $\sim48\,hr$, for $\Delta \vfi\leq 10\,deg$. So if in such case the science session starts with the TM close to an unstable equilibrium, then it will be practically stay in this position for its whole duration.

\subsection{Estimate of the Axial Torque \label{s6.3.}}

By the equation (\ref{eqmot}), we got the expression of the axial torque,
\be
\label{Taxial}
|T_z|=T_*(\tilde z)\,\Delta z\sin^6\left(\Delta \vfi\right)\,\mu(\gamma,\lambda)\,\sin{\gamma}e^{-\tilde{z}^2}\; ,\qquad\mu(\gamma,\lambda)=\frac{1+\lambda^2}
{\left(
1-2\lambda^2\cos{\gamma}+\lambda^4
\right)^2}\; .
\ee
The ballpark number for the torque is obtained by the maximum of the expression (\ref{T*}):
\[
T_*(0)=\frac{\pi^{3/2}}{4}\,\epsilon_0 V_0^2\,\frac{ a}{d}\approx 9.5\times 10^{-14}\,N\;;
\]
of the dimension of a torque per unit length. Contributions of the patch widths, and the patch azimuthal distribution were not taken into account. A more meaningful estimate is obtained by computing the average torque for all the relative positions of the patches. 
Introducing the normalized length $l_0=L/2\Delta z$, we write, using expression (\ref{Taxial}):
\[
|\bar{T}_z|= T_*\Delta z\sin^6
\left(\Delta \vfi\right)\,\frac{1}{l_0}\int_{0}^{l_0}d\tilde{z}e^{-\tilde{z}^2}\;\frac{1}{\pi}\int_0^{\pi}d \gamma\,\mu(\gamma,\lambda)\,\sin{\gamma}\;,
\]
we calculate the two integrals
\[
\int_{0}^{l_0}d\tilde{z}e^{-\tilde{z}^2}=\frac{\sqrt{\pi}}{2}erf(l_0)\;,\qquad
\int_0^{\pi} d \gamma\,\mu(\gamma,\lambda)\,\sin{\gamma}=\frac{2}{\sin^4{\Delta \vfi}\left(1+\cos^2{\Delta \vfi}\right)}\;,
\]
and substitute them in the expression, to find
\[
|\bar{T}_z|=\frac{2}{\sqrt{\pi}}\,T_*\frac{(\Delta z)^2}{L} erf(l_0)\frac{\sin^2
\left(\Delta \vfi\right)}{1+\cos^2{\Delta \vfi}}\,\approx 1.3\times 10^{-12}\,(\Delta z)^2\, erf(l_0)\,\frac{\sin^2
\left(\Delta \vfi\right)}{1+\cos^2{\Delta \vfi}}\,Nm
\;,
\]
with $\Delta z$ in meters. For the typical case $l_0\gg 1$, the above expression becomes
\be
\label{meanTz}
|\bar{T}_z| \approx 1.3\times 10^{-12}(\Delta z)^2\frac{\sin^2\Delta \vfi}{1+\cos^2\Delta \vfi}\,Nm\;,
\ee
which is proportional to the square of the axial patch width; for small angular widths the mean value goes to zero with the square of $\Delta \vfi$, too.

\subsection{Spin Motion for a General Patch Distribution \label{s6.4.}}

Here we make an important concluding remark. It is easy to see that the picture of the spin motion due to the patch effect torque given in section \ref{s6.1.} for the case of two patches is valid, in fact, for any boundary voltage distributions well. Indeed, in any case the electrostatic patch energy is bounded and is a $2\pi$--periodic function of the rotation angle $\gamma$, therefore equations (\ref{eqmotpot})---(\ref{Wpm}) hold, along with the whole following argument about the picture of motion based on them. Thus in the most general case the cylinder either rotates all the way in the same direction, or oscillates about a stable rest point, depending on whether the total energy is above or below the critical value, the global maximum of potential energy. The only significant difference is that for a general patch distribution the number of equilibria can be larger than two (but always even, with the equal number of stable and unstable rest points, since the torque is continuous). 

The increase of the number of equilibrium positions leads to a trend of decreasing the amplitude of the typical oscillatory motion. The remark at the end of section \ref{s6.1.} is also valid in the general case: rotational regimes are not anticipated under the STEP set--up at all. The experiment conditions are such that any friction or other dissipation should be extremely low, so the typical spin oscillations of a TM should be damped but very slowly, as compared even with the duration of a single science session. (One should note, however, that if some resistive energy losses are present, as it appears to have happened with GP-B~\cite{BT}, then this picture might change significantly). These oscillations will make the axial force found in CPEII change periodically with the time. If the basic frequency or some of its harmonics is close to the frequency of the STEP science signal, then it might introduce a systematic error in the experiment, provided that the PE force and the acceleration due to it is large enough. For this reason, as well perhaps for many other ones,  careful pre- and post-mission calibrations on orbit are recommended, along with the extensive simulations before the flight as described in CPEII, section VID.

\begin{acknowledgments}

This work was supported by ICRANet (V.F.) and by KACST through the collaborative agreement with GP-B (A.S.). The authors are grateful to Remo~Ruffini and Francis~Everitt for their permanent interest in and support of this work, and for some valuable remarks. Many our colleagues at GP-B and STEP made valuable remarks and comments on the paper and provided some valuable information. We are greatful to all of them for this, in particular, to Dan DeBra, Sasha Buchman, David Hipkins, John~Mester,  and Paul Worden.
\end{acknowledgments}

\appendix
\section{Calculation of the Slanting Torques}

We provide here intermediate results needed for computing of the slanting torques. Particular terms that appear under the integrals in the formulas (\ref{IntTorquex})---(\ref{pTorquey}) are: 
\be
\label{funclist}
z\frac{\partial \Phi^p}{\partial \rho}\cos{\vfi},\;\,z\frac{\partial \Phi^p}{\partial \rho}\sin{\vfi},\;\,z\frac{\partial \Phi^p}{\partial \vfi}\cos{\vfi},\;\,z\frac{\partial \Phi^p}{\partial \vfi}\sin{\vfi},\;\,\frac{\partial \Phi^p}{\partial z}\cos{\vfi},\;\,\frac{\partial \Phi^p}{\partial z}\sin{\vfi}\;;
\ee
we need to determine the Fourier coefficients of these functions evaluated at the boundary $\rho=b$. 
The radial and the angular derivatives of the potential involved here are: 
\be
\label{AFipder}
\frac{\partial \Fip}{\partial \rho}\Biggl|_{\rho=b}=
-\frac{1}{d}\int_{-\infty}^{\infty}dk\sum_{n=-\infty}^{\infty}e^{\imath\left(kz+n\vfi\right)}
\left[G_n(k)-H_n(k)e^{-\imath\,n\gamma}\right]\;;
\ee
\be
\label{Adphidphi}
\frac{\partial \Phi^p}{\partial \vfi}\Biggl|_{\rho=b}=\frac{1}{2\pi}\,\LLnof e^{i(kz+n\vfi)}\left[in\,H_n(k)e^{-in\gamma}\right]\;;
\ee
they come from the formulas (\ref{Fipder}) and (\ref{dphidphi}), respectively. The $z$ derivative  is computed from the second of the boundary conditions (\ref{bcp}): 
\be
\label{Adphidz}
\frac{\partial \Phi^p}{\partial z}\Biggl|_{\rho=b}=\frac{i}{2\pi}\,\LLnof e^{i(kz+n\vfi)}\left[k\,H_n(k)e^{-in\gamma}\right]\;.
\ee

For any function $u(\vfi,z)$ such that $z\,u(\vfi,z)$ is squarely integrable, by the definition of the Fourier transform (\ref{Four.}), the following equalities hold:
\bea
\label{zFour}
u(\vfi,z)\,z=\frac{i}{2\pi}\int_{-\infty}^\infty \sum_{n=-\infty}^{\infty} e^{i\left(kz+n\vfi\right)}\frac{\partial u_n(k)}{\partial k} \;;\qquad\qquad\qquad\quad\;\;\,\\
u(\vfi,z)\cos{\vfi}=\frac{1}{4\pi}\int_{-\infty}^\infty \sum_{n=-\infty}^{\infty} e^{i\left(kz+n\vfi\right)}\left[u_{n-1}(k)+u_{n+1}(k)\right]\;;\;\;\nonumber\\
u(\vfi,z)\sin{\vfi}=-\frac{i}{4\pi}\int_{-\infty}^\infty \sum_{n=-\infty}^{\infty} e^{i\left(kz+n\vfi\right)}\left[u_{n-1}(k)-u_{n+1}(k)\right]\;.\nonumber
\eea
Using formulas (\ref{zFour}) and (\ref{AFipder}) with the radial derivative playing the role of $u(\vfi,z)$, the first two functions (\ref{funclist}) are represented in the following way:
\bea
\label{dphidrhophicos}
z\frac{\partial \Phi^p}{\partial \rho}\cos{\vfi}\Biggl|_{\rho=b}=\frac{i}{4\pi d}\int_{-\infty}^\infty dk\sum_{n=-\infty}^{\infty} e^{i\left(kz+n\vfi\right)}\frac{\partial }{\partial k}\left[\left(G_{n-1}(k)-H_{n-1}(k)e^{-i(n-1)\gamma}\right)+\right.\\
\left.
\left(G_{n+1}(k)-H_{n+1}(k)e^{-i(n+1)\gamma}\right)\right]\;;\nonumber\\
\label{dphidrhophisin}
z\frac{\partial \Phi^p}{\partial \rho}\sin{\vfi}\Biggl|_{\rho=b}=-\frac{1}{4\pi d}\int_{-\infty}^\infty dk\sum_{n=-\infty}^{\infty} e^{i\left(kz+n\vfi\right)}\frac{\partial }{\partial k}\left[\left(G_{n-1}(k)-H_{n-1}(k)e^{-i(n-1)\gamma}\right)-\right.\\
\left.
\left(G_{n+1}(k)-H_{n+1}(k)e^{-i(n+1)\gamma}\right)\right]\;.\nonumber
\eea

By the same token, using formula (\ref{Adphidphi}) instead of (\ref{AFipder}), for the two terms proportional to the potential derivative in $\vfi$ we obtain:
\bea
\label{dphidphizcos}
z\frac{\partial \Phi^p}{\partial \vfi}\cos{\vfi}\Biggl|_{\rho=b}=-\frac{1}{4\pi}\int_{-\infty}^\infty dk\sum_{n=-\infty}^{\infty} e^{i\left(kz+n\vfi\right)}\frac{\partial }{\partial k}\left[(n-1)H_{n-1}(k)e^{-i(n-1)\gamma}+\right.\\
\left.
(n+1)H_{n+1}(k)e^{-i(n+1)\gamma}\right]\;;\nonumber\\
\label{dphidphizsin}
z\frac{\partial \Phi^p}{\partial \vfi}\sin{\vfi}\Biggl|_{\rho=b}=\frac{i}{4\pi}\int_{-\infty}^\infty dk\sum_{n=-\infty}^{\infty} e^{i\left(kz+n\vfi\right)}\frac{\partial }{\partial k}\left[(n-1)H_{n-1}(k)e^{-i(n-1)\gamma}-\right.\\
\left.
(n+1)H_{n+1}(k)e^{-i(n+1)\gamma}\right]\;.\nonumber
\eea 
Note that  expressions (\ref{dphidrhophicos})---(\ref{dphidphizsin}) here are valid under the additional conditions
\[
\label{addcond1Four}
\int_{-\infty}^\infty dk\sum_{n=-\infty}^{\infty} \Biggl|\frac{\partial }{\partial k} G_{n}(k)\Biggr|^2<\infty\;,\qquad\int_{-\infty}^\infty dk\sum_{n=-\infty}^{\infty} \Biggl|\frac{\partial }{\partial k} H_{n}(k)\Biggr|^2<\infty\;,
\]
which are equivalent to:
\be
\label{addcond1}
\int_{-\infty}^\infty dz \int_{0}^{2\pi} d\vfi \Biggl|zG(\vfi,z)\Biggr|^2<\infty\;;\qquad\int_{-\infty}^\infty dz\int_{0}^{2\pi} d\vfi  \Biggl|zH(\vfi,z)\Biggr|^2<\infty\;.
\ee

To determine the remaining two terms in (\ref{funclist}) containing the derivative of the potential with respect to $z$, we just need the last two formulas from (\ref{zFour}) and the expression (\ref{Adphidz}) in place of $u(\vfi,z)$. This results in: 
\bea
\label{dphidzphicos}
\frac{\partial \Phi^p}{\partial z}\cos{\vfi}\Biggl|_{\rho=b}=-\frac{i}{4\pi}\,\LLnof e^{i(kz+n\vfi)}\,k \left[H_{n-1}(k)e^{-i(n-1)\gamma}+H_{n+1}(k)e^{-i(n+1)\gamma}\right]\;;\quad \\
\label{dphidzphisin}
\frac{\partial \Phi^p}{\partial z}\sin{\vfi}\Biggl|_{\rho=b}=\frac{1}{4\pi}\,\LLnof e^{i(kz+n\vfi)}\,k \left[H_{n-1}(k)e^{-i(n-1)\gamma}-H_{n+1}(k)e^{-i(n+1)\gamma}\right]\;.\qquad
\eea

Formulas (\ref{dphidrhophicos})---(\ref{dphidzphisin}) provide the expressions and the needed conditions (\ref{addcond1}) for calculating the integrals in the formulas (\ref{IntTorquex})---(\ref{pTorquey}) for the slanting torque.

\section{Calculation of the Torque for a Single Patch at each of the Cylinders}

To get the torque we first need to find the Fourier coefficients of the boundary distributions  $G(\vfi,z)$ and $H(\vfi-\gamma,z)$ from the formulas (\ref{Onebcp.}) and their derivative with respect to $k$. The former can be represented by the Fourier coefficients (\ref{VpatchFour.}), (\ref{uonn.}) as:
\be
\label{OnebcpFour.}
G_n(k)=\cV_ n(k,\lambda_1,\Delta z_1)\;;\qquad H_n(k)=\cV_n(k,\lambda_2,\Delta z_2)\;.
\ee

The latter are the derivatives in $k$ of these expressions, and they are given by the general formula:
\be
\label{GnHnder}
\frac{\partial \cV_n^j(k)}{\partial k}=-V_j\Delta z\left(k(\Delta z)^2+iz_j\right)\exp{\left[-\left(\frac{k\Delta z_j}{\sqrt{2}}\right)^2\right]}u_n(\lambda_j)\,e^{-i(kz_j +n\vfi_j)}\;;\qquad j=1,2\;,
\ee
with $\cV_n^1(k)=G_n(k)$ and $\cV_n^2(k)=H_n(k)$.

The torque due to the interaction between the patches and uniform potential difference is a linear function of the derivatives in $k$ of $G_n(k)$ and $H_n(k)$ calculated at $k=0$ and $n=1$. These are determined by the formulas (\ref{GnHnder}) as:
\[
\frac{\partial G_1(k)}{\partial k}\Biggl|_{k=0}=-\sqrt{2\pi}V_1\Delta z_1\left(iz_1\right)\frac{1-\lambda_1^2}{4}e^{-i\vfi_1}\;;\quad\frac{\partial G_1(k)}{\partial k}\Biggl|_{k=0}=-\sqrt{2\pi}V_2\Delta z_2\left(iz_2\right)\frac{1-\lambda_2^2}{4}e^{-i\vfi_2}\;,
\] 
where we used equality (\ref{uonn.}) to express the coefficient $u_n$.
In order to calculate the torque we need just to substitute the above expressions in the formulas (\ref{TxInt}) and (\ref{TyInt}). This leads to the following result:
\bea
\label{oneIntT}
T_x^{Int}=-\sqrt{2\pi^{3}}\,\frac{\epsilon_0 a}{d^2} \Vm\Biggl[V_1\Delta z_1 \left(\sin{\Delta\vfi_1}\right)^2 z_1\sin{\vfi_1} -V_2\Delta z_2 \left(\sin{\Delta\vfi_2}\right)^2 z_2\sin{\left(\vfi_2+\gamma\right)}\Biggr]\;;\\
T_y^{Int}=\sqrt{2\pi^{3}}\,\frac{\epsilon_0 a}{d^2} \Vm\Biggl[V_1\Delta z_1 \left(\sin{\Delta\vfi_1}\right)^2 z_1\cos{\vfi_1} -V_2\Delta z_2 \left(\sin{\Delta\vfi_2}\right)^2 z_2\cos{\left(\vfi_2+\gamma\right)}\Biggr]\;.\nonumber
\eea
Note that here we used the first of the relations (\ref{Deltafi.}) to replace $\lambda$ with the more meaningful parameter $\Delta \vfi$ .

The expressions for the slanting torque due to the interaction between the patches are more cumbersome to find, since one needs to calculate the sum over $n$ and the integral over $k$  of the product of the Fourier coefficients of the boundary distributions and their derivatives in $k$. For the $x$ component, we combine formula (\ref{Txp}) with (\ref{GnHnder}) for the derivatives, and (\ref{OnebcpFour.}) for the boundary functions, to obtain the expression:
\be
\label{OneTxp}
T_x^p=2\pi\, \frac{\epsilon_0 a}{d^2}\int_{-\infty}^\infty dk\, \Re\Biggl\{
-V_1^2\Delta z_1^2{N}(\lambda_1)\left[k \Delta z_1^2+iz_1\right]\exp{\left[- k^2\Delta z_1^2\right]}e^{-i\vfi_1}+
\ee
\[
V_1V_2\Delta z_1\Delta z_2{M}_1\Biggl[\left(k \Delta z_2^2+iz_2\right)e^{ik\left(z_1-z_2\right)}+\left(k\Delta z_1^2+iz_1\right)e^{-ik\left(z_1-z_2\right)}\Biggr]\exp{\left[- \frac{k^2}{2}\left(\Delta z_1^2+\Delta z_2^2\right)\right]}-
\]
\[
V_2^2\Delta z_2^2{N}(\lambda_2)\left[k \Delta z_2^2+iz_2\right]\exp{\left[- k^2\Delta z_2^2\right]}e^{-i\vfi_2}\Biggr\}\;;
\]
for the matter of space we set $\gamma=0$, without any loss of generality, and denoted [compare with CPEII, Appendix A]:
\be
\label{Msdef.}
M_1={M}_1(\lambda_1,\vfi_1,\lambda_2,\vfi_2)\equiv\frac{1}{{2\pi}}\sum_{n=-\infty}^\infty\,u_n(\lambda_1)e^{\imath n\vfi_1}\,u_{n+1}(\lambda_2)e^{-\imath (n+1)\vfi_2}\;;
\ee
\be
N(\lambda)\equiv\frac{1}{{2\pi}}\sum_{n=-\infty}^\infty\,u_n(\lambda)\,u_{n+1}(\lambda)=M_1(0,\lambda;0,\lambda;)\;.
\label{Nsdef.}
\ee
The values of these coefficients are obtained by summing up geometrical progressions [see coefficient $u_n$ in formula (\ref{uonn.})]:
\bea
\label{MN}
M_1=\frac{\left(1-\lambda_1\right)\left(1-\lambda_2\right)}{8}\left\{e^{-\imath \vfi_1}(1+\lambda_1)\left[1+\frac{\lambda_1}{2}(1+\lambda_2)\frac{e^{-\imath(\vfi_1-\vfi_2)}-\lambda_1\lambda_2}{D}\right]+\right.\qquad\qquad\quad\;\\
\left.
e^{-\imath \vfi_2}(1+\lambda_2)\left[1+\frac{\lambda_2}{2}(1+\lambda_1)\frac{e^{\imath(\vfi_1-\vfi_2)}-\lambda_1\lambda_2}{D}\right]
\right\}\;;\;\nonumber D=1-2\lambda_1\lambda_2\cos(\vfi_1-\vfi_2)+\left(\lambda_1\lambda_2\right)^2;\\
N(\lambda)=\frac{1-\lambda^2}{8}(2-\lambda)\;.\qquad\qquad\qquad\qquad\qquad\qquad\qquad\qquad\,
\eea
All we need now to get $T_x^p$ is the two integrals in the formula (\ref{OneTxp}), which are well known:
\[
\int_{-\infty}^\infty dk \exp\left[-\frac{k^2\left(\Delta z_1^2+\Delta z_2^2\right)}{2}\right]e^{\pm\imath k(z_1-z_2)}=\frac{\sqrt{2\pi}}{\sqrt{\Delta z_1^2+\Delta z_2^2}}\exp\left[-\frac{(z_1-z_2)^2}{2\left(\Delta z_1^2+\Delta z_2^2\right)}\right]\; ;
\]
\[
\int_{-\infty}^\infty dk \exp\left[-\frac{k^2\left(\Delta z_1^2+\Delta z_2^2\right)}{2}\right]\,k\,e^{\pm\imath k(z_1-z_2)}=
\pm \imath \sqrt{2\pi}\frac{z_1-z_2}{\left(\Delta z_1^2+\Delta z_2^2\right)^{{3}/{2}}}\exp\left[-\frac{(z_1-z_2)^2}{2\left(\Delta z_1^2+\Delta z_2^2\right)}\right]\;.
\]
In the case $z_1=z_2$ and $\Delta z_1=\Delta z_2=\Delta z$ they becomes:
\[
\int_{-\infty}^\infty dk \exp\left[-k^2\Delta z^2\right]=\sqrt{\pi}/\Delta z\;;\qquad \int_{-\infty}^\infty dk \exp\left[-k^2\Delta z^2\right]\,k=0\;.
\]

With these results we are now able to rewrite formula (\ref{OneTxp}) for the torque as:
\bea
\label{Txpr}
T_x^p=-2\pi^{3/2}\, \frac{\epsilon_0 a}{d^2}\Biggl\{
V_1^2z_1\Delta z_1{N}(\lambda_1)\sin{\vfi_1}+V_2^2z_2\Delta z_2{N}(\lambda_2)\sin{\vfi_2}-\qquad\\
V_1V_2\,\bal\,\Im\left({M_1}\right)
\left[
\sqrt{2}\tilde{z}\left(\Delta z_2^2-\Delta z_1^2\right)+\left(z_1+z_2\right)
\right]\exp{[-\tilde{z}^{\,2}]}\Biggr\}\nonumber\;,
\eea
with the new notations
\be
\label{bartildez}
\bar{l}\equiv\sqrt{2}\frac{\Delta z_1\Delta z_2}{\sqrt{\Delta z_2^2+\Delta z_1^2}}\;;\qquad\qquad \tilde{z}\equiv\frac{z_1-z_2}{\sqrt{2\left(\Delta z_2^2+\Delta z_1^2\right)}}\;.
\ee  

The calculation of the $y$ component of the patch torque does not present any additional difficulties. In the same way as we derived formula (\ref{Txpr}), one can find for $T^p_y$:
\bea
\label{Typr}
T_y^p=2\pi^{3/2}\, \frac{\epsilon_0 a}{d^2}\Biggl\{
V_1^2z_1\Delta z_1{N}(\lambda_1)\cos{\vfi_1}+V_2^2z_2\Delta z_2{N}(\lambda_2)\cos{\vfi_2}-\qquad\\
V_1V_2\,\bal\,\Re\left({M_1}\right)
\left[
\sqrt{2}\tilde{z}\left(\Delta z_2^2-\Delta z_1^2\right)+\left(z_1+z_2\right)
\right]\exp{[-\tilde{z}^{\,2}]}\Biggr\}\nonumber\;,
\eea
[again, the notations (\ref{MN}) and (\ref{bartildez}) are used].

To obtain the closed form representation of the axial torque we combine the formula (\ref{Tz}) with the expressions (\ref{OnebcpFour.}). The integral in $k$ there is found above, so the result is:
\be
\label{OneTzimpl}
T_z^p=4\pi^{3/2}\, \frac{\epsilon_0 a}{d}
V_1V_2\,\bal\,\Im\left(M_3\right)\exp{[-\tilde{z}^{\,2}]}\; ,
\ee
with the coefficient $M_3$ given by
\be
\label{M3def}
M_3=M_3(\lambda_1,\vfi_1,\lambda_2,\vfi_2)\equiv\frac{1}{{2\pi}}\sum_{n=-\infty}^\infty\,n\,u_n(\lambda_1)e^{-\imath n\vfi_1}\,u_{n}(\lambda_2)e^{\imath n\vfi_2}\;.
\ee
Formula (\ref{uonn.}) for $u_n$ leads to an explicit sum of this series reduced to the derivative of a geometric progression:
\be
\label{M3}
M_3=-i\Biggl[\frac{\left(1-\lambda_1^2\right)\left(1-\lambda_2^2\right)}{8}\Biggr]
\frac{\left(1-\lambda_1^2\lambda_2^2\right)}{D^2}
\sin\left(\vfi_1-\vfi_2\right)\;.
\ee
Using this we write formula (\ref{OneTzimpl}) in the final explicit form:
\be
\label{OneTz}
T_z^p=-\frac{\pi^{3/2}}{4}\, \frac{\epsilon_0 a}{d}
V_1V_2\,\bal\,\left(1-\lambda_1^2\right)\left(1-\lambda_2^2\right)\sin\left(\vfi_1-\vfi_2\right)\left(\frac{1-\lambda_1^2\lambda_2^2}{D^2}\right)\,\exp{[-\tilde{z}^{\,2}]}\;.
\ee

\vfill\eject

\vskip20mm
\vfill\eject

\vskip20mm
\vfill\eject

\vskip20mm


\begin{thebibliography}{99}
\bibitem{foot1}This formula offers a method of calculation of the patch effect force alternative to the energy method used in CPEI. We have checked that results of both in the case of unshifted cylinders do coincide.
\end{thebibliography}
\end{document}